\documentclass[twocolumn,amsmath,amssymb]{revtex4}
\usepackage{graphicx}
\usepackage{dcolumn}
\usepackage{bm}

\begin{document}

\title{Quantum Hall ferromagnets and transport properties of buckled Dirac
materials}
\author{Wenchen Luo}
\affiliation{Departement of Physics and Astronomy, University of Manitoba, Winnipeg,
Canada R3T 2N2}
\author{Tapash Chakraborty }
\affiliation{Departement of Physics and Astronomy, University of Manitoba, Winnipeg,
Canada R3T 2N2}
\pacs{xxxx}

\begin{abstract}
We study the ground states and low-energy excitations of a generic Dirac
material with spin-orbit coupling and a buckling structure in the presence
of a perpendicular magnetic field. The ground states can be classified into 
three types under different conditions: SU(2), easy-plane, and Ising quantum 
Hall ferromagnets. For the SU(2) and the easy-plane quantum Hall ferromagnets
there are goldstone modes in the collective excitations, while all the modes
are gapped in an Ising-type ground state. We compare the Ising quantum Hall
ferromagnet with that of bilayer graphene and present the domain wall
solution at finite temperatures. We then specify the phase transitions and
transport gaps in silicene in Landau levels 0 and 1. The phase diagram
strongly depends on the magnetic field and the dielectric constant. We note
that there exists triple points in the phase diagrams in Landau level $N=1$
that could be observed in experiments.
\end{abstract}

\date{\today }
\maketitle

Recently, several graphene-like systems such as silicene, germanene \cite%
{silicene} and transition metal dichalcogenides (MoS$_2^{}$) \cite{wang}
have received considerable attention. They all have a honeycomb geometry in
the $xy$ plane as in graphene \cite{graphene_book,abergeletal}, but
additionally with a buckled structure in the $z$ direction. The buckling is
induced by the atoms that are heavier than the carbon atoms in graphene. The
heavy atoms also have more complex electron orbitals. In these systems the
spin-orbit (SO) coupling is important, while the SO coupling is negligible
in graphene. Hence, the electrons in these systems must be described by a
massive Dirac equation in which the mass is induced by the SO coupling. We
generically call these materials as buckled Dirac materials. The Brillouin
zone is similar to that of graphene: a hexagon with two inequivalent valleys 
$K$ and $K^{\prime}$. In MoS$_2^{}$ the $\Gamma $ point is important because
the energy near this point is close to that in the $K,K^{\prime}$ valleys in
the valence band. For simplicity, in this paper, we consider only the $K$
and $K^{\prime }$ valleys in order to compare with graphene.

In the presence of an applied perpendicular magnetic field the electron bands 
split into a series of Landau levels (LLs) near the two valleys. The fractional 
quantum Hall (QH) effect \cite{fqhe} has been studied recently in silicene and 
germanene \cite{vadym}. The fractal butterflies have also been investigated
theoretically in these systems \cite{fractal}. Here we report on the ground
states and transport properties of the symmetry broken states in the integer
QH effect regime. In bilayer and multilayer graphene \cite{luo1,luo2},
earlier theoretical works have indicated that the ground states in the $
N\neq 0$ LLs are Ising quantum Hall ferromagnets (QHFs), since the
interlayer Coulomb potential is different from the intralayer one. The
resulting transport properties of bilayer graphene were also observed in an
experiment \cite{kayoung}. In buckled Dirac materials the buckling divides
the system into two \textquotedblleft pseudo-layers\textquotedblright. Atoms 
$A$ and $B$ belong to different pseudo-layers, respectively. Hence the
buckling structure makes these monolayer Dirac materials similar to bilayer
graphene, i.e., we could observe the Ising QHF in these one-atom-layer
systems. We discuss below the various QH states and the collective modes of
different QHFs in a few LLs.

The buckling also couples to an external electric field. Without the
magnetic field, silicene and germane may be converted to topological
insulators in an appropriate electric field \cite{tahir}. In the QH regime the
phases and transport properties are also much richer and more interesting
when the electric field is applied. In this work, we will specifically
discuss how the electric field and the dielectric constant (of different
substrates) change the phase diagram and control the spin and valley
pseudo-spin in silicene. These materials are therefore potential candidates
for application in spintronics.

We first consider a generic monolayer Dirac material with the SO coupling.
The Brillouin zone is in general a regular hexagon (as in graphene) with two
inequivalent valleys $K$ and $K^{\prime }$. The noninteracting Hamiltonian
around valley $\eta =K,K^{\prime }$ in a magnetic field is 
\begin{equation}
H_{\eta}^{}=v_F^{}\left(\sigma_x^{}P_x^{}-\eta\sigma_y^{}P_y^{}\right)-
\lambda_{SO}^{} \sigma_z^{},  \label{generalhamiltonian}
\end{equation}
where $\eta=1,-1$ for the $K$ and $K^{\prime}$ valley respectively, $v_F^{}$
is the Fermi velocity, $\sigma$ is the Pauli matrix, $\mathbf{P}=\mathbf{p}+e
\mathbf{A}$ is the canonical momentum and $\lambda_{SO}^{}$ describes the SO
coupling parameter. The SO strength is also described as the mass of the
Dirac fermion near each valley. We choose the Landau gauge of the vector
potential $\mathbf{A}=\left(0,Bx,0\right)$. The LL energy spectrum is given
by $E_0^{}=\pm \lambda_{SO}^{}$ and $E_{n\neq 0}^{}=\mathtt{sgn}\left(
n\right) \sqrt{ \lambda_{SO}^2+2\left( v_F^{}\hbar /\ell\right)^2\sqrt{
n^2+\left\vert n\right\vert}},$ where $\ell=\sqrt{\hbar/eB}$ is the magnetic
length and $n$ is the LL index. The eigen wavefunctions in the two valleys
are 
\begin{equation}
\psi_{n,X}^{K}=\binom{\widetilde{a}h_{n,X}^{}}{\widetilde{b}h_{n-1,X}^{}}
,\psi_{n,X}^{K^{ \prime}}=\binom{\widetilde{b}h_{n-1,X}^{}}{\widetilde{a}
h_{n,X}^{}},  \label{wf}
\end{equation}
where $X$ is the guiding center, $h_{n,X}^{}$ ($h_{n<0}^{}=0$) is the LL
wave function of a two-dimensional electron gas (2DEG) in a conventional
semiconductor. We define $a=\left\vert \widetilde{a}\right\vert,b=|
\widetilde{b}|,$ so the normalization condition is $a^{2}+b^{2}=1$. In MoS$
_{2}^{}$ the $\Gamma$ point should be included in the Hamiltonian. The
low-energy effective Hamiltonian is due to the three \textit{d}-orbitals of
Mo atoms which are located in the same plane \cite{mos2}. So the MoS$_2^{}$
is equivalent to a monolayer system without buckling.

If the geometry of the monolayer Dirac material is exactly the same as
graphene then the valley pseudo-spin has a SU(2) symmetry in any LL.
However, we need to consider the buckling when we calculate the Coulomb
interaction. The ground states of bilayer or multi-layer graphene in the $N\neq
0 $ LL are valley pseudo-spin Ising QHFs \cite{luo1,luo2}. The SU(2)
symmetry of valley pseudo-spin is broken to a Z$_2^{}$ symmetry because
there is a factor $e^{-qd}$ difference between the inter-layer and
intra-layer Coulomb potentials, where $q$ is the momentum and $d$ is the
distance between two layers. We follow the formalism in \cite{luo1,luo2} to
present a more general classification of the QHFs of the ground states in a
buckled Dirac material. We assume a buckling $d$ in the $z$ direction
between the two elements of the wavefunction spinors in Eq.~(\ref{wf}). The
buckling divides the wavefunctions into two pseudo-layers. The density
matrix $\rho$ in the momentum space is 
\begin{equation}
\rho_{\sigma,\sigma^{\prime}}\left(\mathbf{q}\right)=\frac1{N_{\phi}^{}}
\sum_{X_1^{}, X_2^{}}^{}e^{-\frac
i2q_x\left(X_1^{}+X_2^{}\right)}\delta_{X_1^{},X_2^{}+q_y^{}\ell^2}^{}
c_{\sigma,X_1}^{\dagger}c_{\sigma^{\prime},X_2^{}}^{},  \label{rho}
\end{equation}
where $\sigma,\sigma^{\prime}=1\rightarrow \left(K,\uparrow
\right),2\rightarrow \left( K,\downarrow \right),3\rightarrow \left(
K^{\prime},\uparrow \right),4\rightarrow \left( K^{\prime},\downarrow \right)
$ are the valley-spin indices, the LL degeneracy is $N^{}_{\phi},$ and the
creation and annihilation operators of electrons are $c^{\dag},c$. The
average values of the elements of the density matrix fully describe the
system with the Hamiltonian in the Hartree-Fock approximation (HFA) \cite
{supp}.

For a quarter filling of a LL which is equivalent to the case of
three-quarter filling due to the electron-hole symmetry, the ground state
satisfies $\langle\rho_{1,1}^{} \left( 0\right)\rangle+\left\langle
\rho_{3,3}^{}\left( 0\right) \right\rangle =1$. The energy of the liquid
phase, without a constant, is given by $E=2Q\langle \rho_{1,1}^{} \left(
0\right) \rangle \left( \langle \rho_{1,1}^{}\left( 0\right) \rangle
-1\right) e^2/\kappa \ell$, where $\kappa$ is dielectric constant, and 
\begin{equation}
{\tilde Q}\left( \mathbf{q}\right) =\left( a^2-b^2\right)^2d/\ell -X_{K,K}^{}\left( 
\mathbf{q} \right)+X_{K,K^{\prime}}^{}\left( \mathbf{q}\right),
\label{criteria}
\end{equation}
and $Q\equiv {\tilde Q}\left( \mathbf{q=0}\right) $. The first term in Eq.~(\ref
{criteria}) is the capacitive energy. The Fock interaction functions $
X_{K,K}^{}\left(\mathbf{q}\right)$ and $X_{K,K^{\prime}}^{}\left(\mathbf{q}
\right)$ are given by Eq.~(3) in \cite{supp}. We can classify the ground
states as follows. If $Q=0$, the ground state is a SU(2) QHF, since $\langle
\rho_{1,1}^{}\left(0\right)\rangle$ and $\langle \rho_{3,3}^{}
\left(0\right)\rangle$ could be of any value to minimize the energy. If $Q<0$,
the ground state is an Ising QHF. The energy is minimized when $%
\langle\rho_{1,1}^{}\left( 0\right)\rangle$ is either 0 or 1. Finally, if $
Q>0$ the ground state is an easy-plane QHF, i.e., the energy is minimum only
when $\langle \rho_{1,1}^{}\left(0\right) \rangle =\langle \rho_{3,3}^{}
\left( 0\right)\rangle=1/2$.

We define the two-particle Green's function 
\begin{eqnarray*}
&&\chi_{\sigma,\sigma^{\prime},\gamma,\gamma^{\prime}}^{}\left( \mathbf{\ q,q
}^{\prime};\tau_1^{}-\tau_3^{}\right)=N_{\phi}^{}\langle
\rho_{\sigma,\sigma^{\prime}}^{}\left(\mathbf{q}\right)\rangle \langle
\rho_{\gamma,\gamma^{\prime}}^{}\left( -\mathbf{q}^{\prime}\right) \rangle \\
&&-N_{\phi}^{}\langle T\rho_{\sigma,\sigma^{\prime}}^{}\left( \mathbf{q,}%
\tau_1^{}\right) \rho_{\gamma,\gamma^{\prime}}^{}\left( -\mathbf{q}%
^{\prime}, \tau_3^{}\right)\rangle,
\end{eqnarray*}
where $T$\ is the time order operator to study the collective behavior of
the system. We employ the generalized random phase approximation (GRPA) to
solve the equation of motion of the two-particle Green's function. Details
can be found in \cite{cote,jules}. The non-zero lowest-energy collective
mode which is given by the poles of the retarded Green's functions is $
C\left( \mathbf{q\rightarrow 0}\right) =\frac{e^2}{\kappa \ell}\left\vert Q
\left[ \left\langle \rho_{1,1}^{}\left( 0\right)\right\rangle -\left\langle
\rho_{3,3}^{}\left( 0\right) \right\rangle\right] \right\vert,$ which for
small $q$, $C\left( \mathbf{q}\right)\sim q^2$. This collective mode, which
is similar to that of an easy-plane QHF in a double quantum well system
without tunnelling, is a precess mode of the valley pseudo-spin in the $xy$
plane \cite{ezawa,cote}. In the three types of QHFs, when $\mathbf{
q\rightarrow 0,}$ the collective modes are distinguished by their gaps. In a
SU(2) QHF, $Q=0$, so $C\left( \mathbf{q\rightarrow 0}\right)=0.$ It is a
goldstone mode. In an easy-plane QHF, $\langle \rho_{1,1}^{}\left( 0\right)
\rangle-\langle \rho_{3,3}^{}\left( 0\right) \rangle=0,$ so $C\left( \mathbf{
q\rightarrow 0} \right)=0$, which means a goldstone mode still exists. In an
Ising QHF, $Q\neq 0$ and $\left\vert \langle \rho_{1,1}^{}\left( 0\right)
\rangle -\langle \rho_{3,3}^{}\left( 0 \right) \rangle \right\vert =1$. The
goldstone mode disappears and all modes are gapped.

We now consider the two actual materials, silicene and germanene, in a
perpendicular electric field $E_z^{}$. The electric field can control the
phases and the spin polarization, useful for application in spintronics or
valley pseudo-spintronics. The low-energy noninteracting Hamiltonian, in the
basis $\left\{ A\uparrow,B\uparrow,A\downarrow,B\downarrow \right\}$ is \cite
{zeyuan} 
\begin{equation}
H_{\eta}^{}=v_F^{}\left( p_x^{}\tau_x^{}-\eta p_y^{}\tau_y^{}\right)+\eta
\tau_z^{}h+dE_z^{} \tau_z^{}/2,  \label{hamiltonian}
\end{equation}
where $h=-\lambda_{SO}^{}\sigma_z^{}-a_0^{}\lambda_R^{}\left(
p_y^{}\sigma_x^{}-p_x^{} \sigma_y^{}\right),$ $\eta=1$ for valley $K$ and $-1
$ for the $K^{\prime}$ valley, $\tau$ and $\sigma$ are the Pauli matrices
corresponding to the sublattices and the spin, $a_0^{}$ is the lattice
constant, $\lambda_R^{}$ is the Rashba SO (RSO) coupling and the buckling is 
$d$. For silicene (germanene) \cite{liu,parameter}, these parameters are $
v_F^{}=5.5\times 10^5$ $(5.09\times 10^5)$ m/s, $a_0^{}=3.86$ $(4.06)$\AA , $
\lambda_{SO}^{}=3.9$ $(43)$meV, $\lambda_R^{}=0.7$ $(10.7)$meV and $d=0.46$ $
(0.66)$\AA .

The Hamiltonian in Eq. (\ref{hamiltonian}) is more complex than that in Eq.~(
\ref{generalhamiltonian}) since the electric field and the RSO coupling are
included. In the QH the wave function in valley $\alpha$ and orbital $o$ is $
\left( c_{o,1}^{\alpha}h_{o+\alpha_1^{}}^{}\ c_{o,2}^{\alpha}h_{o+\alpha
_2^{}}^{}\ c_{o,3}^{\alpha }h_{o+\alpha _3^{}}^{}\
c_{o,4}^{\alpha}h_{o+\alpha_4^{}}^{} \right)^T,$ with the normalization
condition $\sum_{i=1}^4\left\vert c_{o,i}^{\alpha}\right\vert^2=1,$ where $
K_1^{}=K_4^{}=0,K_2^{}=-1,K_3^{}=1;K_1^{\prime}=-1,K_2^{\prime}=K_3^{
\prime}= 0,K_4^{\prime}=1.$ Because of the RSO coupling the eigenvectors are
not spin polarized. We introduce another degree of freedom, the orbital, to
replace the spin. We find that without Zeeman coupling the energies of the
orbitals $o=N,N-1$ are close to each other in the LL $N$. The concept of the
orbital degree of freedom here is similar to that in the $N=0$ LL in bilayer
graphene. The RSO interaction couples different spins in a valley. For zero
RSO coupling the orbital degree of freedom is identical to spin. In reality, 
$\left\vert c_{N-1,1}^{\alpha}\right\vert,\left\vert
c_{N-1,2}^{\alpha}\right\vert,\left\vert
c_{N,3}^{\alpha}\right\vert,\left\vert c_{N,4}^{\alpha}\right\vert \lesssim
10^{-4}$, and so approximately, the orbital $N$ is associated with spin up
and orbital $N-1$ is associated with spin down. Note that the coefficients $
c_{o,i}^{\alpha}$ not only depend on the magnetic field but also on the
electric field.

We neglect the LL mixing since the LL gap is large ($E_{N=0}^{}\sim
0,E_{N=\pm 1}^{} \approx \pm 60$ meV for silicene). The density matrix and
the many-body Hamiltonian in the HFA are given in \cite{supp}. We define the
Green's function $G_{\alpha,o;\beta,o^{\prime}}^{} \left( X,X^{\prime},\tau
\right)=-\langle Tc_{\alpha,o,X}^{}\left( \tau \right)
c_{\beta,o^{\prime},X^{\prime}}^{\dagger}\left( 0\right) \rangle,$ with a
relation to the density matrix at zero temperature, $G_{\alpha,o;\beta,o^{%
\prime}}^{}\left( \mathbf{q,} \tau=0^{-}\right) =\langle
\rho_{\beta,o^{\prime};\alpha,o}^{}\left(\mathbf{q}\right)\rangle.$ Solving
the equation of motion of the Green's function we could obtain the ground
states of the system \cite{cote2}. In what follows, we define the
valley-orbital (or say, valley-spin if $\lambda_R^{}=0$) indices as $\left(
K,N\right)\rightarrow 1,\left( K,N-1\right)\rightarrow 2,
\left(K^{\prime},N\right) \rightarrow 3,\left( K^{\prime},N-1\right)
\rightarrow 4,$ for simplicity.

Due to the electron-hole symmetry, we consider only the quarter- and
half-filled LLs. The system can be described in the (pseudo-)spin language.
The valley pseudo-spin field in orbital $o$ is defined by $%
p_{o,x}^{}+ip_{o,y}^{}= \left\langle
\rho_{K,o;K^{\prime},o}^{}\right\rangle,p_{o,z}^{}=\left\langle
\rho_{K,o;K,o}^{}\right\rangle-\left\langle
\rho_{K^{\prime},o;K^{\prime},o}^{}\right
\rangle,$ and $\mathbf{p}%
=\sum_o^{}\mathbf{p}_o^{}.$ We could approximately associate the orbital
with the real spin: $o=N$ associated with spin up and $o=N-1$ associated
with spin down. Hence, we define the spin field: $S_{\alpha,x}^{}+iS_{%
\alpha,y}^{}=\left\langle \rho_{\alpha,N;\alpha,N-1}^{}\right\rangle
,S_{\alpha,z}^{}=\left\langle \rho_{\alpha,N; \alpha,N}^{}\right\rangle
-\left\langle \rho_{\alpha,N-1;\alpha,N-1}^{}\right\rangle,$ and $\mathbf{S}%
=\sum_{\alpha}^{}\mathbf{S}_{\alpha}^{}.$

\begin{figure}[tbp]
\includegraphics[width=6.0cm]{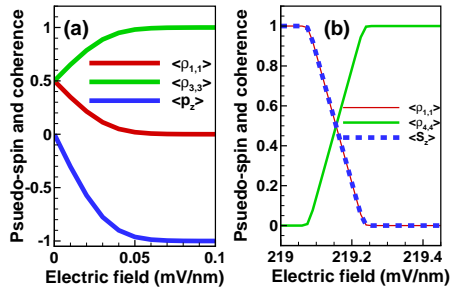}
\caption{(Color online) (a) The occupation of the states $\left\vert
1\right\rangle$ and $\left\vert 3\right\rangle $, and the valley
polarization $\left\langle p^{}_z\right\rangle $ at $\protect\nu =-1$. (b)
Order parameters around the phase transition region at $\protect\nu =0$. }
\label{ll0_rho}
\end{figure}

\textit{Filling factor }$\nu =-1:$ When $E_z^{}=0$, the ground state is an
easy-plane QHF in valley pseudo-spin. It can also be obtained by the
classification parameter $Q$ in Eq.~(\ref{criteria}): $Q>0$ if we
approximate $\lambda_R^{}=0$. The ground state lies on the easy-plane QHF
regime. It is the symmetric state of the two valleys $\left\vert
GS\right\rangle =\left( \left\vert 1\right\rangle+\left\vert 3\right\rangle
\right)/\sqrt2$. When the electric field increases, there is a bias between
the two states $\left\vert 1\right
\rangle $ and $\left\vert 3\right\rangle$
so that the ground state is a bonding state $\left\vert GS\right\rangle
=a_1^{}\left\vert 1\right\rangle+a_3^{}\left\vert 3\right\rangle$. When the
electric field is strong $E_z^{}\gtrsim 0.08$ mV/nm, the bias is large
enough to polarize the valley pseudo-spin. The order parameters in the phase
transition are shown in Fig.~\ref{ll0_rho} (a).

For an easy-plane QHF, the charged excitation is a bimeron or an
anti-bimeron described in the anisotropic nonlinear $\sigma $ model \cite%
{moon,Ezawa_book}. At $E_z^{}=0$ the Lagrangian of this model can be written as $%
L=\frac12\sum_{\mu =x,y,z}^{}\rho_{\mu}^{}\left( \mathbf{\partial}%
_{\mu}^{}m_{\mu}^{}\right)^2,$ where we define the normalized field $\mathbf{%
m=}4\pi\ell^2\mathbf{p.}$ The pseudo-spin stiffnesses $\rho_{\mu}^{}$ are
defined in Eqs.~(25) and (26) in \cite{supp}. The excitation energy of a
bimeron-antibimeron pair is then given by $\delta E_{pair}=8\pi \left(
\rho_x+\rho_y+\rho_z\right) /3$ \cite{moon,Ezawa_book}. For $B=10$T, $E_z^{}=0$, $%
\kappa=1,$ the excitation energy of a bimeron-antibimeron pair is $39.2$
meV. In comparison, the excitation energy of an electron-hole pair is $156$
meV. Hence, the transport gap is due to the bimeron-antibimeron pair.

\textit{Filling factor }$\nu =0:$\textit{\ }For $E_z^{}=0$, $\left\vert
1\right\rangle$ and $\left\vert 3\right\rangle$ are fully occupied, the
ground state is spin polarized and is valley unpolarized. The ground state
is stable when $E_z^{}<219.05$ mV/nm. In the region $E_z^{}\in \left[
219.05,219.25\right]$ mV/nm, the coherence $\left\langle
\rho^{}_{1,4}\right
\rangle$ between $\left\vert 1\right\rangle $ and $%
\left\vert 4\right\rangle$ arises. In Fig.~\ref{ll0_rho} (b), we show the
phase transition region where $\left\langle \rho^{}_{1,1} \right\rangle $
and $\left\langle \rho^{}_{4,4}\right\rangle$ are gradually changed. The
spin of the system is also controlled unpolarized gradually by the electric
field. When $E_z>219.25$ mV/nm the system is spin unpolarized but valley
pseudo-spin polarized, all electrons are in valley $K^{\prime }$.

In the $N>0$ LLs the nature of the broken symmetry states are different from
those in the $N=0$ LL. For $N>0$ the ground state of a LL is an Ising QHF,
which is similar to bilayer graphene \cite{luo1,luo2}. For simplicity and
without loss of generality, we only study the filling factors $\nu =3,4$,
since the LL mixing is important in higher LLs.

\begin{figure}[tbp]
\includegraphics[width=6.0cm]{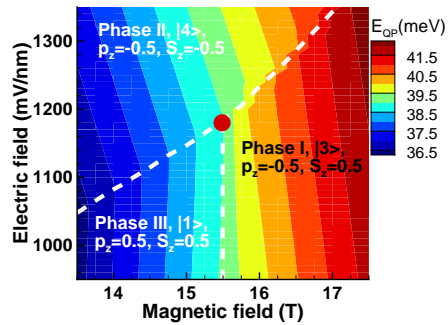}
\caption{(Color online) The excitation energy contour of the quasi-particle
around the triple point at $\protect\nu =3,\protect\kappa =5$.  The dashed
lines are located at the phase transitions. The red dot is the triple point.}
\label{qp_nu3_kappa5}
\end{figure}

\textit{Filling factor }$\nu =3:$ For $E_{z}^{{}}=0$, we assume that $%
\lambda _{R}^{{}}=0$ so that $Q<0.$ Hence the ground state is an Ising QHF
with a valley $Z_{2}^{{}}$ symmetry, which is also supported by our
numerical calculation including $\lambda _{R}^{{}}$. The SU(2) valley
symmetry is broken to a $Z_{2}^{{}}$ symmetry. Figure \ref{qp_nu3_kappa5}
shows the phase diagram in a finite electric field for $\kappa =5$. For $%
\lambda _{R}^{{}}=0$ the SU(2) spin symmetry is also broken to a $Z_{2}^{{}}$
symmetry at the phase transition between Phase I (or Phase III) and Phase
II. These symmetry broken states are all induced by the small buckling
geometry. Note that the valley and spin can also be controlled by the
electric or the magnetic field. At the two sides of the phase transition
line in Fig.~\ref{qp_nu3_kappa5}, the valley or spin is reversed.

Interestingly, for $B=15.5$ T and $E_{z}=1180$ mV/nm, there is a triple
point (the red dot in Fig.~\ref{qp_nu3_kappa5})\ in the phase diagram. This
triple point occurs only when $\kappa \gtrsim 3$. When the dielectric
constant is very large the electron gas is close to a noninteracting system
and the triple point would disappear. At a finite electric field the phase
diagram is also changed by the dielectric constant $\kappa $. For $\kappa =1$
the phase III disappears and only other two phases survive when $%
E_{z}^{{}}<2000$ mV/nm.

The Ising QHF here is similar to that in bilayer or multilayer graphene \cite%
{luo2}, but is different from the Ising QHF system with different LLs \cite%
{isingexp} in semiconductor quantum wells. Hence, the lowest charged
excitation may be a skyrmion around $E_z^{}=0$. However, the skyrmion in
this system must be calculated numerically with a microscopic Hamiltonian in
the symmetric gauge.

\begin{figure}[tbp]
\includegraphics[width=6.0cm]{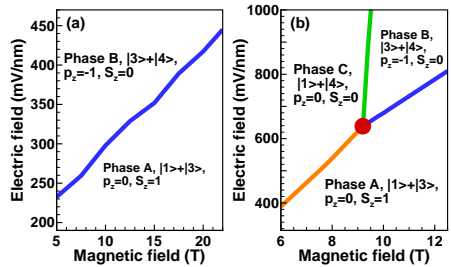}
\caption{(Color online) The phase diagrams at $\protect\nu =4$ and (a) $%
\protect\kappa =1$, and (b) $\protect\kappa =4.$ }
\label{nu4_kappa}
\end{figure}

There is no domain wall at zero temperature. At finite temperature the
domain wall could exist to lower the free energy of the system \cite%
{jungwirth}, when we consider the wall entropy. Below a critical temperature
T$_C$, domain walls provide 1D channels carrying extra charges
(electron-hole pairs) to dissipate the transport charge of the 2DEG, when
the domain walls are dense enough to overlap. So the resistance spike in $%
R_{xx}$ appears. Above T$_C$, the domain wall will be infinitely long and
expand to the sample perimeter. The charge in the domain wall can not
dissipate the transport electrons any more and hence the resistance spike
disappears. Following the study of the domain wall in a graphene bilayer 
\cite{luo2} we obtain the kink domain wall of the valley pseudo-spin at $%
E_z=0$: $m_x\left( \mathbf{r}\right) =\sin \theta \left( \mathbf{r}
\right),m_y=0$ and $m_z\left( \mathbf{r}\right) =\cos \theta \left( \mathbf{r%
}\right)$ with 
\begin{equation}
\theta =2\arctan \exp \left[ \sqrt{2\left(
K^{}_z-K^{}_{\perp}\right)/\rho^{}_s}x\right],  \label{kink}
\end{equation}
where we approximate $\rho^{}_s=\rho^{}_x\approx \rho^{}_z$, and define $%
K^{}_{\perp}= X_{N,N,N,N}^{K,K^{\prime}}\left( \mathbf{0}\right) /(8\pi )$
and $K^{}_z=X_{N,N,N,N}^{K,K}\left( \mathbf{0}\right) /(8\pi).$ The
excitation energy per unit length of the domain wall is then given by $%
\delta E=2\sqrt{2\left( K^{}_z-K^{}_{\perp}\right) \rho^{}_s}$.

We also present the quasi-particle (QP) excitation energy $E^{}_{QP}$ around
the triple point in Fig.~\ref{qp_nu3_kappa5}. Experimentally, the phase
transitions between different spins (Phase I or III to Phase II) can not
only be observed in a NMR experiment, but also in a transport measurement.
The phase transition between different valleys (Phase I to III) may be
observed in a transport or a circular light absorption experiment \cite%
{tabert}. At the phase transitions a resistivity spike may be observed due
to the existence of the domain wall at finite temperature, which has been
reported in a LL Ising QHF \cite{isingexp,jungwirth} and in the Ising QHF of
bilayer graphene \cite{luo2,kayoung}.

\textit{Filling factor }$\nu =4:$ For $\kappa =1,$ the phase diagram in a
magnetic field is shown in Fig.~\ref{nu4_kappa} (a). The triple point
appears when $\kappa \gtrsim 3$. We show an example of the triple point
which is marked as a red dot located at $B=9.2$T, $E_{z}=635$mV/nm for $%
\kappa =4$ in Fig. \ref{nu4_kappa} (b). Phase C sets in when $B<9.2$T, since
the kinetic energy contributes more and the Coulomb interaction $%
e^{2}/\kappa \ell $ is decreased by the large dielectric constant.

\begin{figure}[tbp]
\includegraphics[width=6.0cm]{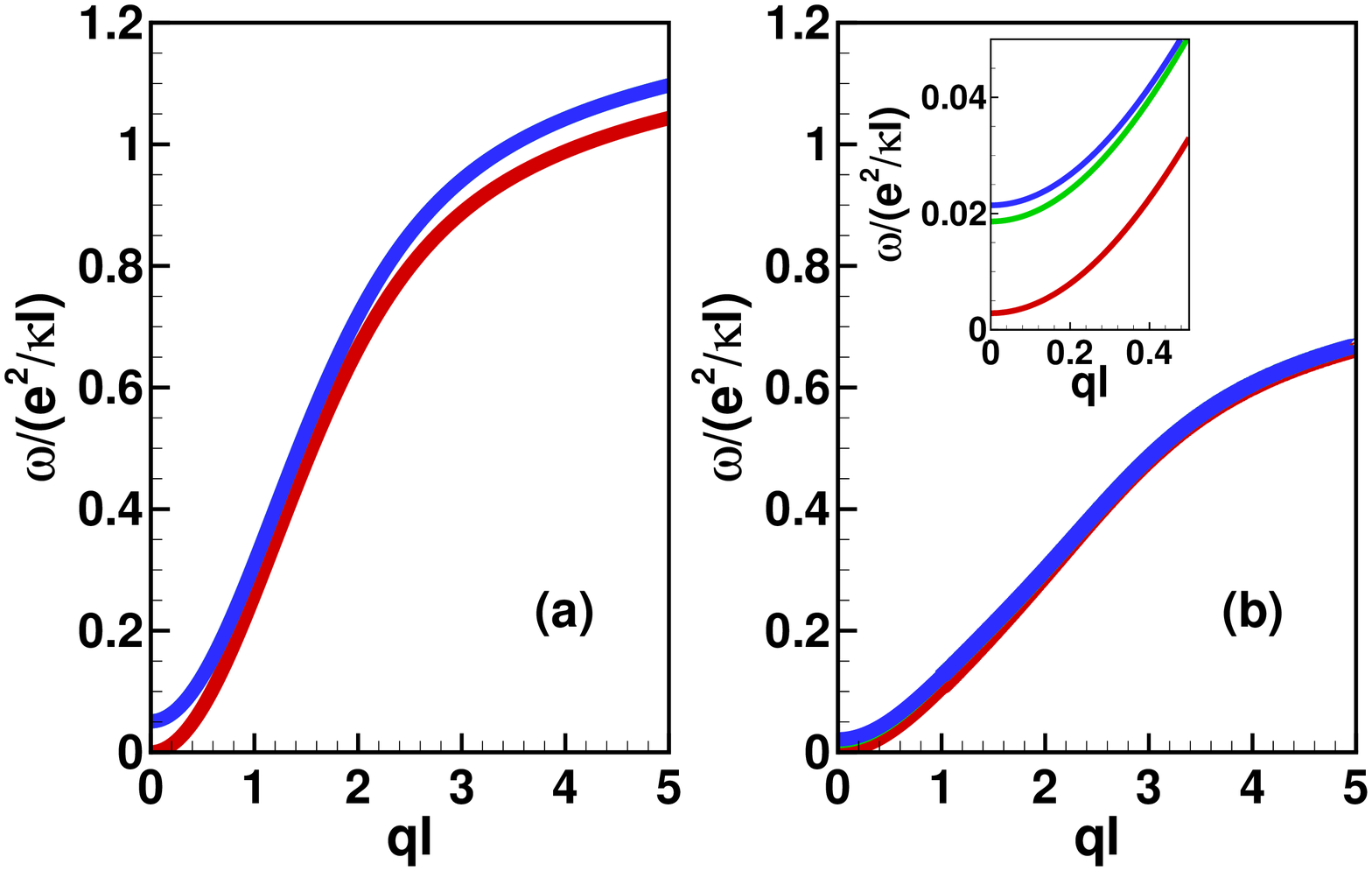}
\caption{(Color online) The collective modes for (a) the easy-plane QHF
ground state at $\protect\nu =-1$ and (b) the Ising QHF ground state at $%
\protect\nu =3,$ when $B=10$T, $\protect\kappa =1$ and $E_z=0.$
The small $q$ region is given as inset.}
\label{cm_nu-1_3}
\end{figure}

As we discussed above, the easy-plane QHFs have a goldstone mode while all
the modes of Ising QHFs are gapped. In silicene we only show the collective
modes at $\nu =-1$ and $3$ in Fig.~\ref{cm_nu-1_3} for simplicity. Note
that, in the region $E^{}_z\in \left[ 219.05,219.25\right] $ mV/nm, the
ground state is a bonding state with a goldstone mode at $\nu =0$.

To summarize, we classify the ground states of a generic buckled Dirac
material in a magnetic field into three types of QHFs. The low-energy
collective modes of the three QHFs are given analytically in the GRPA. A
goldstone mode exists in the SU(2) and in the easy-plane QHFs, but not in
the Ising QHF. We then focus on a real material, viz., silicene. Without an
electric field we note that in silicene the magnetic field is able to change
the coefficients $c_{o,i}^{\alpha }$ in the wavefunctions. However, in a
very small magnetic field ($B\ll 0.01$T) the ground state becomes an
easy-plane QHF at $\nu =3$. In such a low magnetic field the QH effect can
not be realized and the LL mixing is not negligible. If the SO coupling can
be tuned then the coefficients of wave functions can be modified. For $B=10$%
T, $\nu =3,\kappa =1$, we find that when $\lambda _{SO}^{{}}>750$ meV, the
ground state at $\nu =3$ is an easy-plane QHF. For about $\lambda
_{SO}^{{}}=750$ meV, the ground state is a SU(2) QHF. This is also true for
germanene. If we could efficiently tune the wave function we may realize the
phase transition between different QHFs. Experimentally, this transition is
observable: the domain wall induced resistivity spike occurs only in an
Ising QHF. The phase diagrams and transport properties in the $N=0,1$ LLs in
silicene also depend on the magnetic field and the dielectric constant which
dramatically change the Coulomb interaction. We have shown the triple points
in Figs.~\ref{qp_nu3_kappa5} and \ref{nu4_kappa}. The SU(2) symmetry of the
spin and valley are broken by the electric field and the buckling structure.
The phase transitions may indeed be observed in NMR, transport or light
absorption experiments.

The work has been supported by the Canada Research Chairs Program of the
Government of Canada.

\end{document}


\title{SUPPLEMENTAL MATERIAL \\
Quantum Hall ferromagnets and transport properties of buckled Dirac materials}
\author{Wenchen Luo}
\affiliation{Departement of Physics and Astronomy, University of Manitoba, Winnipeg,
Canada R3T 2N2}
\author{Tapash Chakraborty}
\affiliation{Departement of Physics and Astronomy, University of Manitoba, Winnipeg,
Canada R3T 2N2}
\pacs{xxxx}
\date{\today }
\maketitle

\section{Many-body Hamiltonian of generic Dirac materials with bcukling}

In a generic two-dimensional (2D) Dirac material with bcukling, the eigen wave functions are 
given by Eq.~(2) in the letter. We divide the wavefunctions into those for two pseudo-layers 
between which there is a separation $d$. The electrons essentially belong to a double-layer 
system. Using the density matrix defined in Eq.~(3) in the letter, the many-body Hamiltonian 
in the Hartree-Fock approximation (HFA) is given by 
\begin{eqnarray}
H &=&\sum_{\sigma}^{}\left( E_{\sigma}^{}+\frac{e^2}{\kappa\ell}\frac d{2\ell}\sum_{\sigma^{\prime}}^{}
y_{\sigma^{\prime}}^{}\langle \rho_{\sigma^{\prime},\sigma^{\prime}}^{}\left(0\right) \rangle 
y_{\sigma}^{}\right) \rho_{\sigma,\sigma}^{}\left( 0\right)  \nonumber \\
&&+\frac{e^2}{\kappa \ell}\sum_{\sigma,\sigma^{\prime}}^{}\overline{\sum_{\mathbf{q}}^{}}
H_{\eta,\eta^{\prime}}^{}\left(\mathbf{q} \right) \langle \rho_{\sigma,\sigma}^{}\left( 
-\mathbf{q}\right) \rangle \rho_{\sigma^{\prime},\sigma^{\prime}}^{}\left( \mathbf{q}
\right)   \nonumber \\
&&-\frac{e^2}{\kappa\ell}\sum_{\sigma,\sigma^{\prime}}^{}\sum_{\mathbf{q}}^{}X_{\eta,\eta^{\prime}}^{}
\left( \mathbf{q}\right) \langle \rho_{\sigma,\sigma^{\prime}}^{}\left( -\mathbf{q}\right)
\rangle \rho_{\sigma^{\prime},\sigma}^{}\left( \mathbf{q}\right),  \label{h}
\end{eqnarray}
where $\kappa$ is the dielectric constant, $\sigma,\sigma^{\prime}=1\rightarrow \left(K,\uparrow \right),
2\rightarrow \left( K,\downarrow \right),3\rightarrow \left( K^{\prime},\uparrow \right),4\rightarrow
\left( K^{\prime},\downarrow \right)$ are the valley-spin indices, $\eta$ and $\eta^{\prime}$ are the 
valley indices in $\sigma $ and $\sigma^{\prime}$ respectively, and $y_1^{}=y_2^{}=a^2,y_3^{}=
y_4^{}=-b^2.$ $E_{\sigma }^{}$ is the kinetic energy of level $\sigma$ with the Zeeman coupling. The
summation with a bar excludes the term of $\mathbf{q=0.}$ The functions $H_{\eta,\eta^{\prime}}^{}$ 
and $X_{\eta,\eta^{\prime}}^{}$ describe the Hartree and Fock interactions between valleys $\eta$ and 
$\eta^{\prime}$, 
\begin{eqnarray}
H_{\eta,\eta^{\prime}}^{}\left( \mathbf{q}\right) &=&\frac1{q\ell} \xi_{\eta,\eta^{\prime}}^{}
\left( q\ell\right), \\
X_{\eta,\eta^{\prime}}^{}\left( \mathbf{q}\right) &=&\int_0^{\infty}dp J_0^{}\left( pq\ell\right) 
\xi_{\eta,\eta^{\prime}}^{}\left(p\right),
\end{eqnarray}
where $J$ is the Bessel function and 
\begin{eqnarray}
\xi_{\eta,\eta}^{} &=&a^4f_{n,n}^{}+b^4f_{n-1,n-1}^{}+2a^2b^2f_{n,n-1}^{}e^{-qd/\ell }, \\
\xi_{\eta,\overline{\eta }}^{} &=&\left(a^4f_{n,n}^{}+b^4f_{n-1,n-1}^{}\right) e^{-qd/\ell}+
2a^2b^2f_{n,n-1}^{},
\end{eqnarray}
with $\overline{\eta}$ being the valley other than $\eta.$ The function $f$ is 
\begin{equation}
f_{n,m}^{}\left( q\right) =e^{-q^2/2}L_n^{}\left( q^{2}/2\right) L_m^{}\left( q^2/2\right),
\end{equation}
with a Laguerre polynomial $L_n^{}$ $\left( L_{n<0}^{}=0\right)$.

For a liquid phase at quarter filling of a Landau level the ground state is polarized. The nonzero 
density matrix elements give the distribution of electrons $\langle \rho_{1,1}^{}\left( 0\right) 
\rangle,\langle \rho_{3,3}^{}\left( 0\right) \rangle$ and the coherence $\langle \rho_{1,3}^{}
\left( 0\right)\rangle,\langle \rho^{}_{3,1}\left(0\right)\rangle$ 
\begin{eqnarray}
\langle \rho_{1,1}^{}\left( 0\right)\rangle+\langle \rho_{3,3}^{}\left( 0\right) \rangle  &=&1, \\
\langle \rho_{1,3}^{}\left( 0\right)\rangle &=&\langle \rho^{}_{3,1}\left( 0\right) \rangle =
\sqrt{\langle \rho^{}_{1,1}\left( 0\right) \rangle \langle \rho^{}_{3,3}\left( 0\right)\rangle},
\end{eqnarray}
where we choose all order parameters to be real. Hence, the energy is given by
\begin{eqnarray}
E &=&\sum_{\sigma=1,3}^{}\left( E_{\sigma }^{}+\frac{e^2}{\kappa\ell}\frac d{\ell}\sum_{\sigma^{\prime}
=1,3}^{}y_{\sigma^{\prime}}^{}\langle \rho_{\sigma^{\prime},\sigma^{\prime}}^{}\left(
0\right) \rangle y_{\sigma}^{}\right)\langle \rho_{\sigma,\sigma}^{}\left( 0\right) \rangle   \nonumber \\
&&-\frac{e^2}{\kappa\ell}\sum_{\sigma,\sigma^{\prime}=1,3}^{}X_{\eta,\eta^{\prime}}^{}\left( 
\mathbf{0}\right) \langle \rho_{\sigma,\sigma^{\prime}}^{}\left( 0\right) \rangle 
\rho_{\sigma^{\prime},\sigma}^{}\left( 0\right).
\end{eqnarray}
By using the conditions $X_{K,K^{\prime}}^{}=X^{}_{K^{\prime},K},$ $X^{}_{K,K}=X^{}_{K^{\prime},
K^{\prime}}$ and ignoring the constant kinetic energy, we obtain
\begin{equation}
E=2\frac{e^2}{\kappa\ell}\left[ \left( a^2-b^2\right)^2d/\ell -X_{K,K}^{}\left( \mathbf{0}\right) 
+X_{K,K^{\prime}}^{}\left( \mathbf{0 }\right) \right] \langle \rho_{1,1}^{}\left( 0\right) \rangle 
\left( \langle \rho_{1,1}^{}\left(0\right)\rangle -1\right).
\end{equation}

\section{Many-body Hamiltonian of silicene}

The elements of the density matrix with extra orbital index can be obtained by modifying Eq.~(3) 
in the letter:
\begin{equation}
\rho_{\alpha,o;\beta,\sigma^{\prime}}^{}\left( \mathbf{q}\right) = \frac1{N_{\phi}^{}}
\sum_{X_1^{},X_2^{}}^{}e^{-\frac i2 q_x^{}\left( X_1^{}+X_2^{}\right)}\delta_{X_1^{},X_2^{}+q_{y}^{}
\ell^2}^{}c_{\alpha,o,X_1}^{\dagger}c_{\beta,\sigma^{\prime},X_2^{}}^{},
\end{equation}
where $\alpha,\beta \,$\ are valley indices, $o,o^{\prime}$ are orbital indices. Since the orbitals 
are not conserved the many-body Hamiltonian in the HFA is more complex than that in 
Eq.~(\ref{h}),
\begin{eqnarray}
&&H=\frac{e^2}{\kappa \ell}\sum_{\alpha,o}^{}\widetilde{E}_{\alpha,o}^{}\rho_{\alpha,o;\alpha,o}^{}
\left(0\right)   \nonumber \\
&&+\frac{e^2}{\kappa\ell}\sum_{\alpha,\beta}^{}\sum_{o_1^{},\ldots,o_4^{}}^{}\overline{\sum_{\mathbf{q}}^{}}
H_{o_1^{},o_2^{},o_3^{},o_4^{}}^{\alpha,\beta}\left( \mathbf{q}\right) \langle \rho_{\alpha,o_1^{};\alpha,o_2^{}}^{}
\left( -\mathbf{q}\right) \rangle \rho_{\beta,o_3^{};\beta,o_4^{}}^{}\left( \mathbf{q}\right)  \nonumber \\
&&-\frac{e^2}{\kappa \ell}\sum_{\alpha,\beta}^{}\sum_{o_1^{},\ldots,o_4^{}}^{}\sum_{\mathbf{q}}^{}
X_{o_1^{},o_4^{},o_3^{},o_2^{}}^{\alpha,\beta }\left(\mathbf{q}\right) \langle \rho_{\alpha,o_1^{};
\alpha,o_2^{}}^{}\left( -\mathbf{q}\right) \rangle \rho_{\beta,o_3^{};\beta,o_4^{}}^{}\left( \mathbf{q}\right),
\end{eqnarray}
where $\widetilde{E}_{\alpha,o}^{}$ contains the kinetic energy $E_{\alpha,o}^{}$ of the orbital $o$ in valley 
$\alpha $ and a capacitive energy, 
\begin{eqnarray}
&&\widetilde{E}_{\alpha,o}^{}=E_{\alpha,o}^{}+\frac d{\ell }\left[\frac{\nu}2-\sum_{\beta,o^{\prime}}^{}
U_{\alpha,o;\beta,o^{\prime}}^0\langle\rho_{\beta,o^{\prime};\beta,o^{\prime}}^{}\left(0\right)\rangle\right], \\
&&U_{\alpha,o;\beta,o^{\prime}}^0=\left(\sum_{i=1,3}\sum_{j=2,4}+\sum_{i=2,4}\sum_{j=1,3}\right) \left\vert
c_{o,i}^{\alpha}\right\vert^2\left\vert c_{o^{\prime},j}^{\beta}\right\vert^2,
\end{eqnarray}
with the filling factor $\nu $. The Hartree interaction is 
\begin{eqnarray}
H_{o_1^{},o_2^{},o_3^{},o_4^{}}^{\alpha,\beta}\left( \mathbf{q}\right) 
&=&\sum_{m,n=1}^2\sum_{i=m,m+2}\sum_{j=n,n+2}G_{1,2,3,4}^{\alpha,\beta;i,j}P_{m,n}^{}\left( q\ell\right)   
\nonumber \\
&&\times \frac{F_{1,2,3,4}^{}}{q\ell}\Theta_{o_1^{}+\alpha_i^{},o_2^{}+\alpha_i^{}}^{}\left( q\ell\right) 
\Theta_{o_3^{}+\beta_j^{},o_4^{}+\beta_j^{}}^{}\left( q\ell \right) 
\end{eqnarray}
where we define a function $P_{m,n}^{}\left( q\right)=e^{-\left\vert
m-n\right\vert qd/\ell -q^2/2}$ and coefficients
\begin{eqnarray}
G_{1,2,3,4}^{\alpha,\beta;i,j} &=&c_{o_1^{},i}^{\alpha \ast}c_{o_2^{},i}^{\alpha}c_{o_3^{},j}^{\beta 
\ast}c_{o_4^{},j}^{\beta}, \\
F_{1,2,3,4}^{} &=&e^{-i\left( o_1^{}-o_2^{}\right) \theta}e^{-i\left( o_3^{}-o_4^{}\right) \left(\theta+\pi\right)},
\end{eqnarray}
with the angle $\theta$ between vector $\mathbf{q}$ and $x$ axis. Function $\Theta$ is defined as 
\begin{equation}
\Theta_{k,l}^{}\left( p\right)=\frac{\sqrt{\min \left( k,l\right)!}}{\sqrt{\max\left( k,l\right)!}}
\frac{\left( ip\right)^{\left\vert k-l\right\vert}}{\sqrt2^{\left\vert k-l\right\vert}}L_{\min \left(
k,l\right)}^{\left\vert k-l\right\vert}\left( \frac{p^2}2\right).
\end{equation}
The Fock interaction is 
\begin{eqnarray}
X_{o_1^{},o_4^{},o_3^{},o_2^{}}^{\alpha ,\beta }\left( \mathbf{q}\right) 
&=&\sum_{m,n=1}^2\sum_{i=m,m+2}\sum_{j=n,n+2}G_{1,4,3,2}^{\alpha,\beta;i,j}
\int dpP_{m,n}^{}\left( p\right)   \nonumber \\
&&\times \Theta_{o_1^{}+\alpha_i^{},o_4^{}+\alpha_i^{}}^{}\left( p\right)\Theta_{o_3^{}
+\beta_j^{},o_2^{}+\beta_j^{}}^{}\left(-p\right) J_{o_1-o_4+o_3-o_2}\left( pq\ell\right).
\end{eqnarray}
Once we obtain the HF Hamiltonian, we could obtain the order parameters of
the ground state by solving the equation of motion of the single Green's
function \cite{cote2}.

\section{Pseudo-spin stiffness of the anisotropic Nonlinear $\protect\sigma$ model}

We extract the Fock energy functional from the Hamiltonian, 
\begin{equation}
H_F=-\frac{e^2}{\kappa \ell}\sum_{\alpha,\beta}\sum_{o_1,o_2,o_3,o_4}
\sum_{\mathbf{q}}X_{o_1,o_4,o_3,o_2}^{\alpha,\beta}\left( \mathbf{q}\right)
\left\langle \rho_{\alpha,o_1;\beta,o_2}\left( -\mathbf{q}\right)
\right\rangle \rho_{\beta,o_3;\alpha,o_4}\left(\mathbf{q}\right),
\end{equation}
where we neglect the kinetic energy, the capacitive energy and the Hartree
interaction. This is because the kinetic energy is a constant, capacitive
energy is very small, and the Hartree term is zero in the slow varying
density approximation.

This field theory is only valid when $E^{}_z=0$. If $E^{}_z\neq 0$, then the
ground state is nolonger an easy-plane QHF. The ground state will be valley
polarized very rapidly, i.e. the $E^{}_z$ is very small to polarize the
valley pseudo-spin. In the finite $E^{}_z$ case, the best way is to calculate
the microscopic Hamiltonian in the symmetric gauge.

If we consider the two valleys only in orbital $o=N$, then the orbital index
could be neglected, and 
\begin{equation}
{\cal E}^{}_F=-\frac12\frac{e^2}{\kappa \ell }\sum_{\alpha,\beta}
\sum_{\mathbf{q}}X_{N,N,N,N}^{\alpha,\beta}\left( \mathbf{q}\right) \langle
\rho^{}_{\alpha,\beta}\left( -\mathbf{q}\right)\rangle \langle
\rho^{}_{\beta,\alpha}\left( \mathbf{q}\right) \rangle.
\end{equation}
Then the Lagrangian of the anisotropic Nonlinear $\sigma$ model is obtained
by calculating the excitation energy when the density matrix is slowly
varied \cite{moon},  
\begin{equation}
L=\frac12\sum_{\mu =x,y,z}^{}\rho_{\mu }^{}\left( \mathbf{\partial}_{\mu }^{}
m_{\mu}^{}\right)^2,
\end{equation}
where the normalized pseudo-spin field $\mathbf{m}$ is defiend by 
\begin{eqnarray}
m_x\left( \mathbf{r}\right)+im_y\left( \mathbf{r}\right) &\mathbf{=}
&4\pi \ell^2\langle \rho_{K,K^{\prime}}^{}\left( \mathbf{r}
\right)\rangle, \\
m_z^{}\left(\mathbf{r}\right) &=&4\pi \ell^2\left( \langle
\rho_{K,K}^{}\left(\mathbf{r}\right) \rangle-\langle \rho_{K^{\prime},K^{\prime }}^{}
\left( \mathbf{r}\right) \rangle \right),
\end{eqnarray}
and the pseudo-spin stiffnesses are given by
\begin{eqnarray}
\rho^{}_z &=&-\frac1{8\pi \ell^2}\lim_{q\rightarrow 0}\nabla_q^2X_{N,N,N,N}^{K,K}\left( \mathbf{q}\right), \\
\rho^{}_x &=&\rho^{}_y=-\frac1{8\pi\ell^2}\lim_{q\rightarrow 0}\nabla_q^2X_{N,N,N,N}^{K,K^{\prime}}
\left( \mathbf{q}\right).
\end{eqnarray}
The excitation energy of a bimeron or an anti-bimeron is 
\begin{equation}
\delta E=\frac{4\pi}3\left( \rho^{}_x+\rho^{}_y+\rho^{}_z\right).
\end{equation}
Note that this energy which is obtained in the field theory is not identical
to a single charged excitation of the electron gas. However, the excitation
energy of a bimeron-antibimeron pair in the field theory is identical to
that in the electron gas, which is double of $\delta E,$
\begin{equation}
\delta E_{pair}=\frac{8\pi}3\left( \rho^{}_x+\rho^{}_y+\rho^{}_z\right).
\end{equation}
This energy is related to the transport property of the system, which can be
measured in a transport experiment.